Primitive chain network simulations for double peaks in shear stress under fast flows of bidisperse entangled polymers

*Yuichi Masubuchi,

Department of Materials Physics, Nagoya University,

Nagoya 464-8603, Nagoya Japan

*To whom correspondence should be addressed

e-mail mas@mp.pse.nagoya-u.ac.jp

ABSTRACT

A few experiments have reported that the time development of shear stress under fast start-up shear deformations exhibits double peaks before reaching the steady state for bimodal blends of entangled linear polymers in specific conditions. To analyze the molecular origin of this phenomenon, multi-chain slip-link simulations based on the primitive chain network model were conducted for the data of a bimodal polystyrene solution reported by Osaki et al. [J Pol Sci B Pol Phys, 38, 2043 (2000)] Owing to the reasonable agreement with the data and the simulation results, the stress was decomposed into contributions from long and short-chain components and decoupled into segment number, stretch, and orientation. The analysis revealed that the first and second peaks correspond to the short-chain orientation and the long-chain stretch, respectively. The results also imply that the peak positions are not affected by the mixing of short and long chains, whereas the second peak intensity depends on the mixing conditions in a complicated manner, dominating the emergence of double peaks.

INTRODUCTION

Viscosity growth under high shear of entangled polymers has been widely investigated as a typical example of the non-linear response [1–6]. Due to the elastic response, viscosity linearly grows against applied strain in the initial stage, irrespective of the strain rate. When the shear rate is smaller than the slowest relaxation rate (the reciprocal longest relaxation time), viscosity gradually reaches zero-shear viscosity after the longest relaxation time. Under high shear, the steady state viscosity decreases with increasing the shear rate, exhibiting shear thinning. Before reaching the steady state, viscosity exhibits an overshoot. When the shear rate is lower than the reciprocal Rouse time, the peak is located at the total shear strain of ca. 2.3, reflecting segment orientation[7]. With higher shear, the total strain at the peak increases due to the contribution of chain stretch with increasing the shear rate. In some literature, viscosity exhibits an undershoot following the overshoot, related to the coherent tumbling of molecules[8–11]. Under extremely high shear, strain hardening without a steady state has been reported[12,13].

Concerning the viscosity overshoot, some literature reports multiple peaks rather than the widely observed single peak. Kinouchi et al.[14] reported viscosity growth curves with two peaks for polystyrene solutions with bidisperse molecular weight distributions in which a small amount of long-chain portion was dispersed in short-chain matrices. Osaki et al.[15] performed an extended study that varied the molecular weight of the short-chain matrix. They reported that double-peak behavior is observed only when the molecular weights are sufficiently separated. Snijkers et al.[16] reported multiple peaks for viscosity growth of a comb polymer melt. These multiple overshoots may be attributable to flow instabilities and edge fracture [17–19]. However, Snijkers et al.[16] performed their experiments with a cone-partitioned plate rheometry[19] to exclude possible effects of edge fracture.

The molecular origin of multiple peaks has yet to be discussed. Concerning bidisperse blends of linear polymers, Kinouchi et al.[14] and Osaki et al.[15] speculated that each mixed portion exhibits its peak. Isram[20] theoretically supported such an idea using his tube model. In particular, he concluded that the first and second peaks reflect overshoots in the stretch of short and long chain fractions, respectively. However, his theoretical model needs to be further considered, particularly regarding the effects of constraint release. In the model, the coupling between different fractions is taken into account by the change of chain contraction through convective constraint release; the relaxation time is modified only under fast flow via chain stretch. In such an implementation, the contribution of segment orientation may need to be further considered.

Nevertheless, the comparison to the data by Osaki et al.[15] was not attained straightforwardly, and no discussion of the segment orientation was given. Other molecular theories that can predict non-linear rheology of mixtures [21–23] have not been applied to this problem.

This paper examined the double peaks in viscosity growth under high shear for a bidisperse polystyrene solution reported by Osaki et al.[15] employing a multi-chain slip-link simulation[24,25], where the coupling between different components is directly considered. The results demonstrated that the second peak is due to the long chain stretch, as Isram showed[20]. However, the first peak is attributable to the segment orientation, and the contribution from the short-chain stretch is minor. Details are explained below.

MODEL AND SIMULATIONS

Because the model employed in this study is the same as that used in the previous studies for shear flows[25–32], only a brief description is given below. In the primitive chain network model, a network consisting of nodes, strands, and dangling ends represents an entangled polymeric liquid. Each polymer chain in the system corresponds to a path connecting dangling ends through nodes and strands. This path is conceptually equivalent to the primitive path in the tube model[33]. However, no tube is considered in this model, and the path fluctuates due to force balance at each network node and dangling end. Namely, the positions of nodes and dangling ends obey a Langevin-type equation of motion, in which force balance is considered among the drag force, tension acting on each strand, osmotic force suppressing density fluctuations, and thermal random force. Mimicking entanglements, a slip-link is located at each node to bundle two subchains. The chains slide through slip-links, and the chain sliding is described by the change rate equation of the number of Kuhn segments on each strand according to the same force balance for the node position along the chain backbone. Because of chain sliding, when a dangling end protrudes from the connected slip-link beyond a critical Kuhn segment number, a new node with a slip-link is created by hooking another segment randomly chosen from the surroundings. Vice versa, when a dangling end slides off from the connected slip-link, the slip-link disappears, and the bundled chains are released. Concerning simulations under shear flows, infinitesimal step shear strain is applied affinely to the network nodes and dangling ends, followed by relaxations according to the force balance mentioned above. This set of deformation and relaxation is repeatedly applied to attain a designated shear rate.

The length, energy, and time units are chosen as the average strand length under equilibrium $a$, thermal energy $kT$, and diffusion time of the node $\tau_0 = \zeta a^2/6kT$, where $\zeta$ is the friction coefficient of the node. For convenience in conversion to the experimental system, instead of $a$ and $kT$, units of molecular weight $M_0$ and modulus $G_0$ are used. Here, $M_0$ is the average molecular weight of strands, and $G_0 = kT/a^3$. As discussed earlier, $M_0$ and $G_0$ are similar but different from the entanglement molecular weight $M_e$ and plateau modulus $G_N$. Nevertheless, the parameters $M_0$, $G_0$, and $\tau_0$ are determined according to the following procedure. $M_0$ is estimated from entanglement molecular weight $M_e$ as $M_0 = 2M_e/3$ according to the empirical relation established earlier[34–37]. Using $M_0$, the average strand number per chain under equilibrium $Z_0$ is determined as $Z_0 = M/M_0$, where $M$ is the molecular weight of the examined polymer. A primitive chain network is created according to this $Z_0$, and an equilibrium simulation run is conducted for a sufficiently long period, at least ten times longer than the longest relaxation time of the system. From stress fluctuations under equilibrium, linear relaxation modulus $G(t)$ is calculated with the Green-Kubo formula. The obtained $G(t)$ is converted to $G'(\omega)$ and $G''(\omega)$ through the REPTATE software[38,39]. These calculations are done with dimensionless units, and the obtained moduli are converted to the experimental value via $G_0$, which is determined from $M_0$ by $M_0 = \rho RT/G_0$. Here, $\rho$ is the polymer density. Finally, $\tau_0$ is determined by fitting $G'(\omega)$ and $G''(\omega)$ to experimental data.

The examined system was polystyrene tricresyl phosphate solutions at 0°C with a polymer concentration of 0.11 g/cm³. $M_0$, $G_0$, and $\tau_0$ were chosen at $10^5$ Da, 2.5e³ Pa, and 0.12 sec, respectively. The molecular weights of short and long chains are $M_L = 7.1 \times 10^5$ Da and $M_H = 8.2 \times 10^6$ Da, which were replaced by the chains with the segment number per chain of $Z_L = 7$ and $Z_H = 82$. Hereafter, subscripts L and H denote the low and high molecular weight components; thus, L does not mean the long chain fraction but the short one. According to the experiment[15], the fractions for the long and short chains were 0.09 and 0.91.

The simulations were performed with a cubic simulation box with periodic boundary conditions. The Lees-Edwards boundary was employed for the shear gradient direction for the cases under shear flows. The box size was $16^3$, which is sufficiently larger than the radius of gyration of examined chains under equilibrium. The segment density was fixed at 10. See Fig 1 for a typical snapshot of the single long chain. Because $M_0$ for the examined system is large, finite chain extensibility [40] was not considered. Friction reduction [41,42] was also neglected because the polymer concentration was low. Eight independent simulation runs were conducted for each shear rate, starting from different initial configurations, and quantities reported below are ensemble averages taken among these simulations.

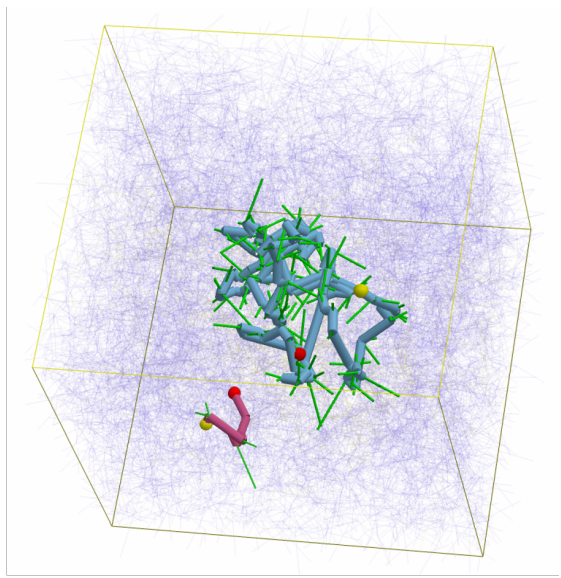

**Figure 1** A typical snapshot of the system. For clarity, consecutive blue and red cylinders show a pair of long and short chains. Red and yellow spheres indicate the chain ends. Bold green lines show the entangled segments for the indicated chains. Thin blue lines draw other segments. The simulation box is displayed as the yellow frame cube.

RESULTS AND DISCUSSION

Figure 2 shows the linear viscoelastic response. The simulation reasonably captures the experimental data with two plateaus in $G'(\omega)$ corresponding to entanglements between all the chains and only between the long chains. However, the longest relaxation time is overestimated. Because the primitive chain network simulation has attained semi-quantitative agreement for bidisperse polystyrene melts in previous studies[43,44], the discrepancy may be due to solvent quality that depends on temperature. Nevertheless, the reason is unknown.

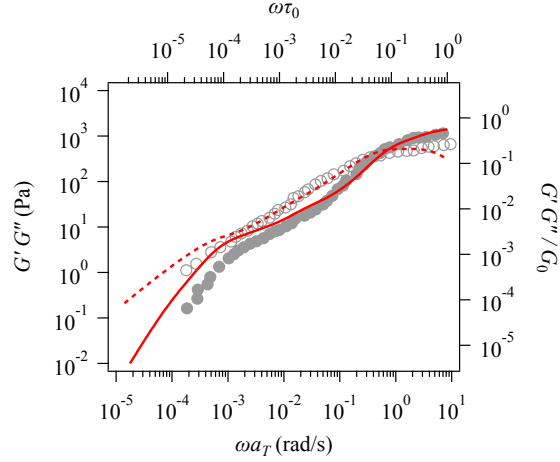

**Figure 2** The linear viscoelasticity of the examined bidisperse polystyrene solution. Symbols represent experimental data from the literature [15] plotted against the left and bottom axes. Red curves show the simulation results.

In Figure 3, panels (a) and (b) show the time development of shear stress sigma and normal stress difference $N_1$ for some shear rates. For low shear rates, both sigma and $N_1$ are quantitatively reproduced by the simulation. As the shear rate increases, the discrepancy between the data and the simulation result worsens. Concerning the shear stress shown in panel (a), the time development up to 10 sec is nicely captured even under high shear. However, after exhibiting the first peak, the second peak in the simulation is weaker and occurs earlier than that observed experimentally. Since the double peaks behavior is not apparent in the simulation even for the highest shear rate reported experimentally, the result for a higher shear is added, as shown by the blue curve, which exhibits double peaks.

After bumping, the shear stress reaches a steady value underestimated in the simulation. This discrepancy in the steady value is summarized in panel (c), which shows the steady-state viscosity plotted against the shear rate. The simulation results (red triangle) are smaller than the data (black circle) in the entire range of examined shear rates. However, the simulation is consistent with the Cox-Merz rule; the steady state viscosity agrees with the complex viscosity (red broken curve). The viscosity obtained experimentally is located beyond the complex viscosity (black broken curve). Nevertheless, the discrepancy in the steady-state viscosity is not significant in the logarithmic plot in panel (c).

The normal stress difference in Fig 3 (b) demonstrates that the data is not correctly reproduced when the shear rate is high. In a short period of up to 20 seconds, the simulation captures the data irrespective of the shear rate. However, the simulation reaches the maximum at an $N_1$ value significantly lower than the experiment. The steady-state value is also underestimated. Note that the blue curve is the simulation result for $\dot{\gamma}=2.5\text{sec}^{-1}$, for which the experiment was not performed, and the coincidence with the data at $\dot{\gamma}=1.165\text{sec}^{-1}$ (uppermost symbols) is meaningless. Nevertheless, $N_1$ exhibits a single peak even under high shear, for which the shear stress shows double peaks. See blue curves in panels (a) and (b).

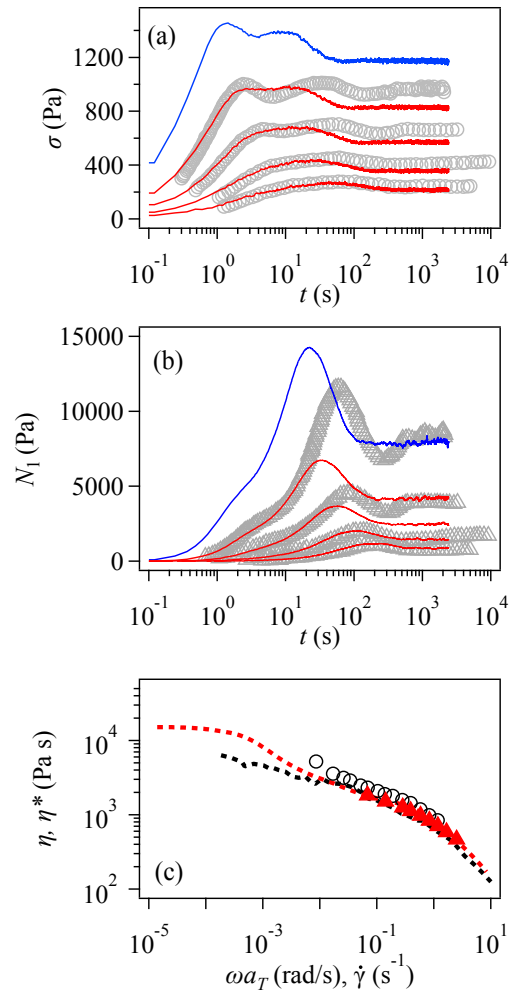

**Figure 3** The time development of shear stress (a) and normal stress difference $N_1$ (b) for the shear rate of 1.165, 0.58, 0.28, and 0.138 $\text{sec}^{-1}$ from Osaki et al. [15] shown by symbols compared with the simulation indicated by red curves. Blue curves are the simulation results for the shear rate of 2.5 $\text{sec}^{-1}$, for which the experimental data are unavailable. Panel (c) shows the steady-state viscosity from Osaki et al. [15] and simulations, indicated by the black and red symbols. Black

and red dotted curves are the complex viscosity from the experiment and simulation.

Although the simulation does not excellently reproduce the data, owing to the reasonable agreement, the origin of the double peaks in shear stress is analyzed below for the case of the blue curve at the shear rate of 2.5sec$^{-1}$ in Figure 3 (a). Figure 4 shows the contributions from the two components for the shear stress and the normal stress difference compared to the response from the entire system (blue) and the results of simulations without mixing. Concerning the double peaks for shear stress seen in panel (a), the first and second peaks come from short (green) and long (red) components, respectively, as suggested by Kinouchi et al.[14] The short-chain contribution shown by the green solid curve coincides with that obtained for the pure short-chain system displayed by the green broken curve, implying that the mixing with the long chain does not affect the short-chain behavior. In contrast, the mixing suppresses the peak stress for the long-chain contribution, whereas the peak position is unchanged. See the red curves. For the first normal stress difference, the peak originates from the long component, whereas the contribution from the short chains appears as a shoulder in the short-time region. The effects of mixing in short and long-chain contributions are similar to that observed for shear stress.

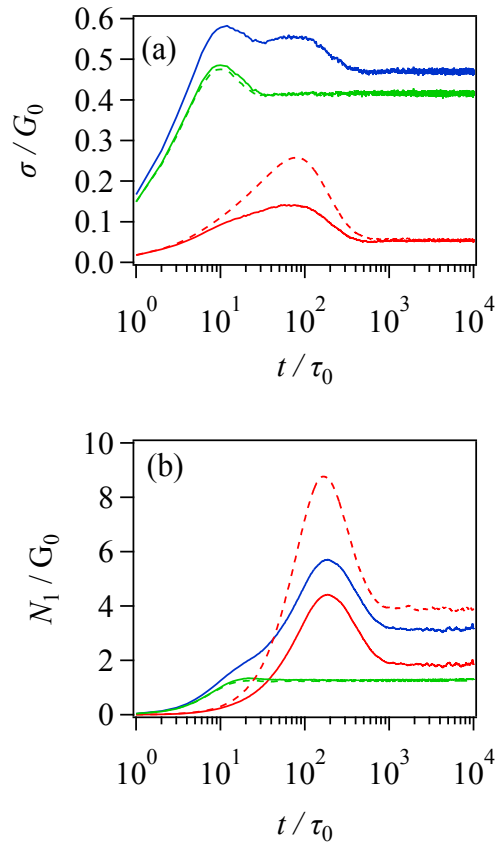

**Figure 4** Shear stress (a) and the first normal stress difference (b) at $\dot{\gamma}\tau_0 = 0.3$. Blue, green, and red curves show the results for the entire system and contributions from short and long chains. Green and red broken curves indicate the results for pure long and short-chain systems without mixing reduced according to their mixing fractions.

Further analysis for the molecular mechanism is shown below according to the decoupling approximation [7,28], in which the stress is approximated as follows. Note that the non-linear spring constant is not considered since the finite chain extensibility is neglected in this study.

$$\boldsymbol{\sigma} \approx 3\left(\frac{Z}{Z_0}\right)\lambda^2 \mathbf{S}$$

Here, $\boldsymbol{\sigma}$ is the stress tensor, $Z$ is the number of segments under deformations, $\lambda^2$ is the squared segment stretch, and $\mathbf{S}$ is the segment orientation tensor. Figure 5 shows these molecular quantities for each component, compared to the cases for the long and short chains without mixing.

Concerning the short chain, all the examined quantities are insensitive to mixing with the long chain, as seen as the coincidence between solid and broken green curves. The peak in shear stress corresponds to the shear component of the orientation tensor $S_{xy}$ shown in panel (c), and it is not related to the stretch $\lambda^2$ in panel (b), according to the peak position. The peak of $S_{xy}$ is located at the applied shear strain of ca. 2.3, consistent with the tube theory, and no peak is observed for $\lambda^2$. The change in $Z$ is relatively minor, slightly decreasing with time. The first normal stress shown in Fig 4 (b) is consistent with the orientation $S_{xx}-S_{yy}$ and $\lambda^2$, both monotonically increase and reach a plateau around $t/\tau_0 \sim 20$.

For the long chain, mixing with the short chain suppresses the magnitudes of $Z$ reduction and $\lambda^2$ enhancement, as seen in panels (a) and (b). However, the characteristic times of their development are unchanged. For instance, the peak position of $\lambda^2$ and the mitigation times of $\lambda^2$ and $Z$ to the steady state are insensitive to mixing. These results are rationalized by the fact that the Rouse relaxation and chain contraction are insensitive to the change in entanglement network. For the orientation $S_{xy}$ shown in panel (c), the peak position is not affected by mixing and is located at the same strain as the short chain. An interesting feature is the undershoot following the peak enhanced by mixing, reflecting the coherent tumbling of the long chains [8,9]. However, the undershoot does not appear in shear stress because the increase in the segment stretch conceals it. Indeed, the position of undershoot in $S_{xy}$ is close to that in $\lambda^2$. For the orientation $S_{xx}-S_{yy}$, no effect of mixing is seen, and thus, the reduction of $N_1$ due to mixing in Fig 4 (b) originates from the reduction of $\lambda^2$ in panel (b).

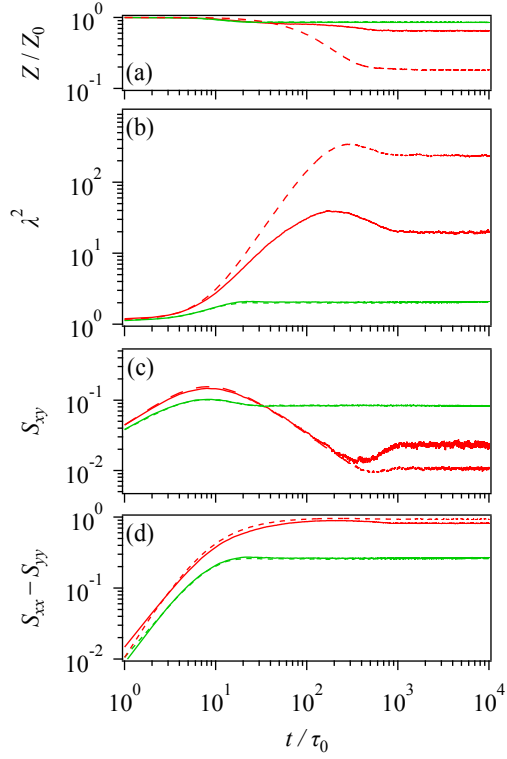

**Figure 5** Decoupled analysis of stress concerning the normalized segment number per chain $Z/Z_0$ (a), squared segment stretch (b), the shear component of orientation tensor (c), and the difference between normal orientation components of orientation tensor (d). Green and red curves indicate behaviors of the short and long chains, respectively. Solid and broken curves display the cases with and without mixing of short and long chains. The applied shear rate is $\dot{\gamma}\tau_0 = 0.3$.

From the analysis above, the mechanism of double peaks in the shear stress of the bidisperse blend can be summarized as follows. Since each peak comes from the response of each component, as shown in Fig 4 (a), the first necessary condition is a separation of the peak positions of the two components. This condition is fulfilled if the Rouse times of two components are well separated, and the applied shear rate is larger than at least one of the reciprocal Rouse times since the peak delays due to the emergence of chain stretch. In the examined case, $\tau_R/\tau_0 = Z_0^2/2\pi^2 \sim$ 5 and 700 for the short and long chains, respectively[45,46]. The normalized reciprocal Rouse times $\tau_0/\tau_R$ are 0.2 and 1.4× $10^{-3}$. These values are not affected by mixing since chain stretch occurs irrespective of entanglement. The applied shear rate is $\dot{\gamma}\tau_0 = 0.3$, which is larger than those values. However, since it is not significantly greater than for the short chain, the short chain stretch only weakly affects the first peak, which is dominated by the segment orientation. In contrast, the second peak originated by the long chain is well delayed due to the long chain stretch. As a result, the two well-separated peaks are realized. In addition to this first condition concerning the peak positions, the second necessary condition is that the magnitude of peaks must be comparable; if the second peak from the long chain component is too large, the first peak is concealed, and vice versa. The long-chain stress is reduced by mixing, naively due to the reduced long-chain density. However, mixing also induces suppression of the long chain stretch by culling out the entanglements between long chains, and the corresponding stress overshoot is diminished. See Fig. 4 (a). For the examined case, due to this suppressed long chain stretch, the second peak is somewhat smeared. Consequently, the balance of molecular weights and volume fractions between two components is difficult. The explanation above is consistent with the earlier conjectures [14,15] and the theoretical analysis [20]. Nevertheless, the behaviors of molecular parameters apart from chain stretch have been shown for the first time.

CONCLUSIONS

The molecular origin of double peaks in the time development of shear stress under fast shear of the bidisperse polystyrene solution reported by Osaki et al. [15] was analyzed with the primitive chain network simulations. The experimental data for linear and non-linear viscoelastic responses under shear were qualitatively reproduced, including the double peaks. Owing to the nature of multi-chain simulation, the entire stress was decomposed into contributions from different components, and the result demonstrated that the first and second peaks come from the short and long-chain components, respectively. Decoupling analysis of stress for each component revealed that the first and second peaks originate from the short-chain orientation and the long-chain stretch, respectively. Although these results are qualitatively consistent with earlier studies, they also imply that the necessary conditions for the emergence of double peaks are not easily fulfilled. Specifically, the condition is not trivial in realizing comparable peak intensities from the balance in molecular weights and volume fractions between long and short-chain components. This complication is probably the reason why earlier reports for the double peaks are only a few, apart from the experimental difficulties. Further experiments and simulations to clarify such a condition are worth conducting, and the results from supplemental studies will be published elsewhere.

ACKNOWLEDGEMENTS

This study was financially supported in part by JST-CREST (JPMJCR1992) and JSPS KAKENHI